\documentclass[amsmath,amssymb,preprint,aps,prl,superscriptaddress,floatfix]{revtex4-1}
\usepackage{graphicx}
\usepackage{color}
\usepackage{epstopdf}
\usepackage{here}
\usepackage{ulem}
\usepackage{sidecap}

\begin{document}

\title{Protection of excited spin states by a superconducting energy gap}
\author{B.\ W.\ Heinrich}
\email{e-mail: bheinrich@physik.fu-berlin.de}
\affiliation{Institut f\"ur Experimentalphysik, Freie Universit\"at Berlin, Arnimallee 14, 14195 Berlin, Germany\\e-mail: bheinrich@physik.fu-berlin.de\\}
\author{L.\ Braun}
\altaffiliation{present address: Fritz-Haber-Institut der Max-Planck-Gesellschaft, 14195 Berlin, Germany}
\affiliation{Institut f\"ur Experimentalphysik, Freie Universit\"at Berlin, Arnimallee 14, 14195 Berlin, Germany\\e-mail: bheinrich@physik.fu-berlin.de\\}
\author{ J.\ I.\ Pascual}
\affiliation{Institut f\"ur Experimentalphysik, Freie Universit\"at Berlin, Arnimallee 14, 14195 Berlin, Germany\\e-mail: bheinrich@physik.fu-berlin.de\\}
\affiliation{CIC nanoGUNE, 20018 Donostia-San Sebasti\'an, Spain}
\affiliation{Ikerbasque, Basque Foundation for Science, 48011 Bilbao, Spain}
\author{K.\ J.\ Franke}
\affiliation{Institut f\"ur Experimentalphysik, Freie Universit\"at Berlin, Arnimallee 14, 14195 Berlin, Germany\\e-mail: bheinrich@physik.fu-berlin.de\\}
\date{\today}



\maketitle 

\textbf{
The latest concepts for quantum computing and data storage envision to address and manipulate single spins.
A limitation for single atoms or molecules in contact to a metal surface are the short lifetime of excited spin states, typically picoseconds, due to the exchange of energy and angular momentum with the itinerant electrons of the substrate~\cite{balashov09PRL,lothSci10,kahleNanoL11,khajetPRL11}.
 Here we show that paramagnetic molecules on a superconducting substrate exhibit excited spin states with a lifetime of $\tau \approx 10$~ns. We ascribe this increase in lifetime by orders of magnitude to the depletion of electronic states within the energy gap at the Fermi level. This prohibits pathways of energy relaxation into the substrate and allows for electrically pumping the magnetic molecule into higher spin states, making superconducting substrates premium candidates for spin manipulation. We further show that the proximity of the scanning tunneling microscope tip modifies the magnetic anisotropy.\\
}

The most efficient way of energy quenching on metallic substrates is the creation of electron-hole pairs~\cite{loth10Nat}. A common strategy to decouple spin states from their electronic environment is to include a non-conductive spacer~\cite{heinrich04, tsukahara09,kahleNanoL11}, with an energy gap in the density of states at the Fermi level  ($E_F$). Superconductors exhibit a tiny but perfect energy gap around the Fermi level due to the condensation of electrons into Cooper pairs at low temperatures. Since the energy scale of superconducting pairing and spin excitations are typically similar, superconductors are ideal materials for stabilizing excited spin states, combining a perfect gap in the density of states (DoS) at $E_F$ with normal metal conductivity, which still allows for addressing the spin by conducting leads. 

\begin{figure}[t]
\includegraphics[width=0.96\textwidth,clip=]{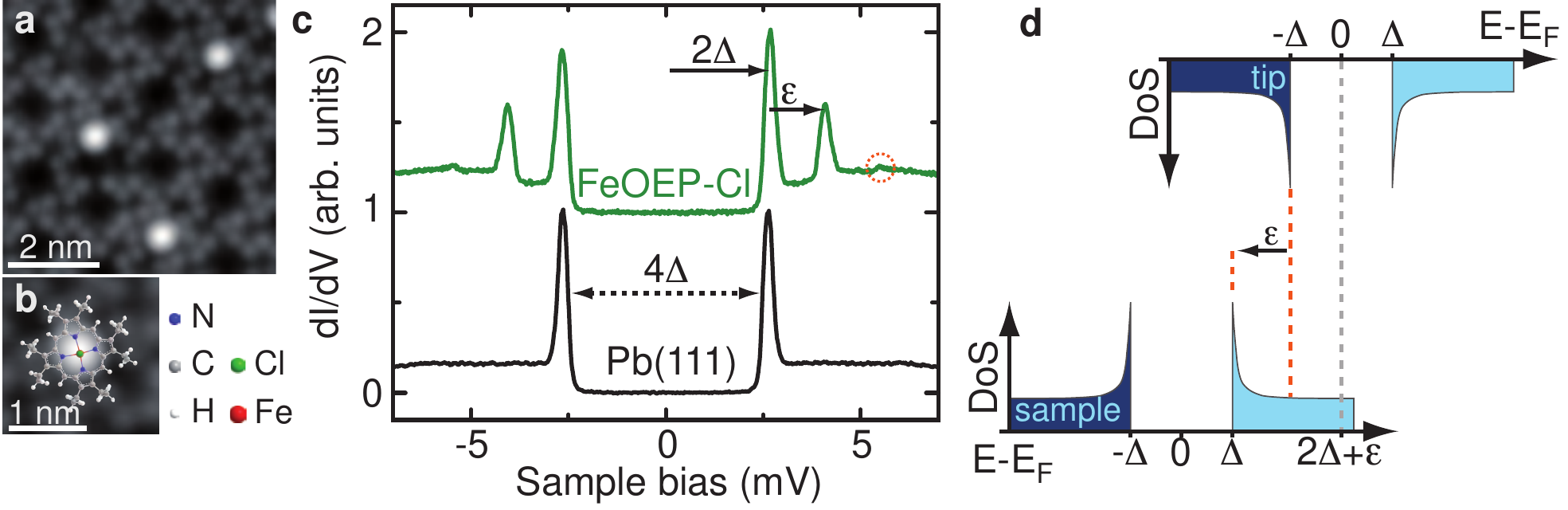}
\caption{Fe-OEP-Cl on Pb(111). (a) STM topography of a mixed island of Fe-OEP-Cl (molecules with a bright protrusion) and Fe-OEP (molecules with a dark center) (Scanning conditions: $V=-50$~mV, $I=200$~pA). (b) Zoom on a single Fe-OEP-Cl with superimposed molecular structure.
(c) $dI/dV(V)$ spectra acquired above pristine Pb(111) and Fe-OEP-Cl (feedback loop opened at $I=200$~pA and $V = 50$~mV followed by an approach $\Delta z = -110$ and $0$~pm, respectively; spectra normalized to unity at the energy of the quasi-particle peaks and offset for clarity). The quasi-particle peaks at $|eV|=2\Delta$ indicate an unperturbed superconducting state. A pair of strong peaks and weak peaks (marked by a dashed circle) in the $dI/dV(V)$ on the molecule are signatures of inelastic tunneling in a superconductor-vacuum-superconductor junction, as sketched in (d): peaks appear at energies $|eV|=2\Delta + \varepsilon$ due to the opening of an inelastic tunneling channel.}
\label{Fig1}
\end{figure}

However, magnetism and superconductivity do not easily coexist. Exchange interaction of a magnetic species with Cooper pairs affects the superconducting ground state and gives rise to new states within the energy gap~\cite{shiba68,yazdani97,ji08,franke11}. To overcome the exchange coupling to the superconductor, our study focuses on paramagnetic metal-organic molecules, whose molecular ligand with inert organic groups separates the central magnetic ion from its conducting environment. At the same time, the organic skeleton provides an anisotropic environment~\cite{tsukahara09}, leading to non-degenerate magnetic eigenstates in the absence of an external magnetic field~\cite{gatteschi06}. We study Fe-octaethylporphyrin-chloride (Fe-OEP-Cl; structure as in Fig.~\ref{Fig1}b) adsorbed on Pb(111), whose Fe center has, in gas phase and bulk, a $+3$ oxidation state with a spin of $S=\frac{5}{2}$ and an in-plane anisotropy (anisotropy parameter $D>0$)~\cite{nishio01,wende07}. 

To address the dynamics of excited spin states of Fe-OEP-Cl on a superconducting Pb substrate, we use a scanning tunneling microscope (STM) at a temperature of 1.2~K with a Pb-covered tip (see \textit{Methods} for details). In the differential conductance [$dI/dV(V)$] spectra on the bare Pb(111) substrate, the depletion of the density of states around $E_F$ is reflected by a gap terminated by two sharp quasi-particle [Bardeen-Cooper-Schrieffer (BCS)] peaks at $|eV|= \Delta_{tip}+\Delta_{sample}=2\Delta= 2.7$~meV, i.e., twice the superconducting gap (see Fig.~\ref{Fig1}c, bottom; $\Delta_{tip}$ ($\Delta_{sample}$) is the pairing energy of the tip (sample),  $e$ the elementary charge).

Fe-OEP-Cl self-assembles on the Pb(111) surface in large quasi-hexagonal molecular islands (Fig.~\ref{Fig1}a). STM images show that many molecules have lost their Cl ligand upon deposition~\cite{bheinrich13}. 
To characterize the spin state of Fe-OEP-Cl, $dI/dV(V)$ spectra were acquired at the center, on top of the Cl ligand (see Fig.~\ref{Fig1}c). The energy gap and quasi-particle resonances of the lead substrate are observed unchanged on the Fe-OEP-Cl molecules, revealing that the superconducting state is unaffected by the presence of the paramagnetic molecule. We thus conclude that, due to the decoupling ligand, no noticeable magnetic interaction  of the molecular spin with the superconductor occurs.

The presence of the paramagnetic molecule causes, instead, remarkable peak features outside the superconducting gap. They appear symmetric with respect to $E_{F}$, indicating an origin related to inelastic electron tunneling phenomena \cite{jaklevic66,stipe98,heinrich04}. In normal metals, the opening of an inelastic tunneling channel due to a molecular excitation produces a step-wise increase of the differential conductance at the excitation energy $\varepsilon$. Here, the effect of the superconducting DoS of tip and sample is twofold: first, it causes a shift of $2\Delta$ in the energy of the inelastic onsets, which appear now at $|eV|=2\Delta+\varepsilon$, and second, it induces a repetition of the BCS peaks at this excitation threshold due to the peaked DoS  (Fig.~\ref{Fig1}d). The energy of the inelastic excitation in the spectra of Fig.~\ref{Fig1}c is $\varepsilon_1=1.4$~meV and the relative amplitude with respect to the BCS peaks is $\approx$ 0.4. These values are typical for spin excitations of paramagnetic species in the presence of ligand field anisotropy \cite{hirji07,hirji06,tsukahara09}. The inelastic tunneling occurs when electrons exchange energy and angular momentum of $\Delta M_S=\pm1, 0$~\cite{hirji06, hirji07} -- $M_S$ are the spin eigenstates -- with the spin of the Fe$^{3+}$ center and induce the non-equilibrium population of excited states in the spin multiplet. 

\begin{figure}[t]
\includegraphics[width=0.96\textwidth,clip=]{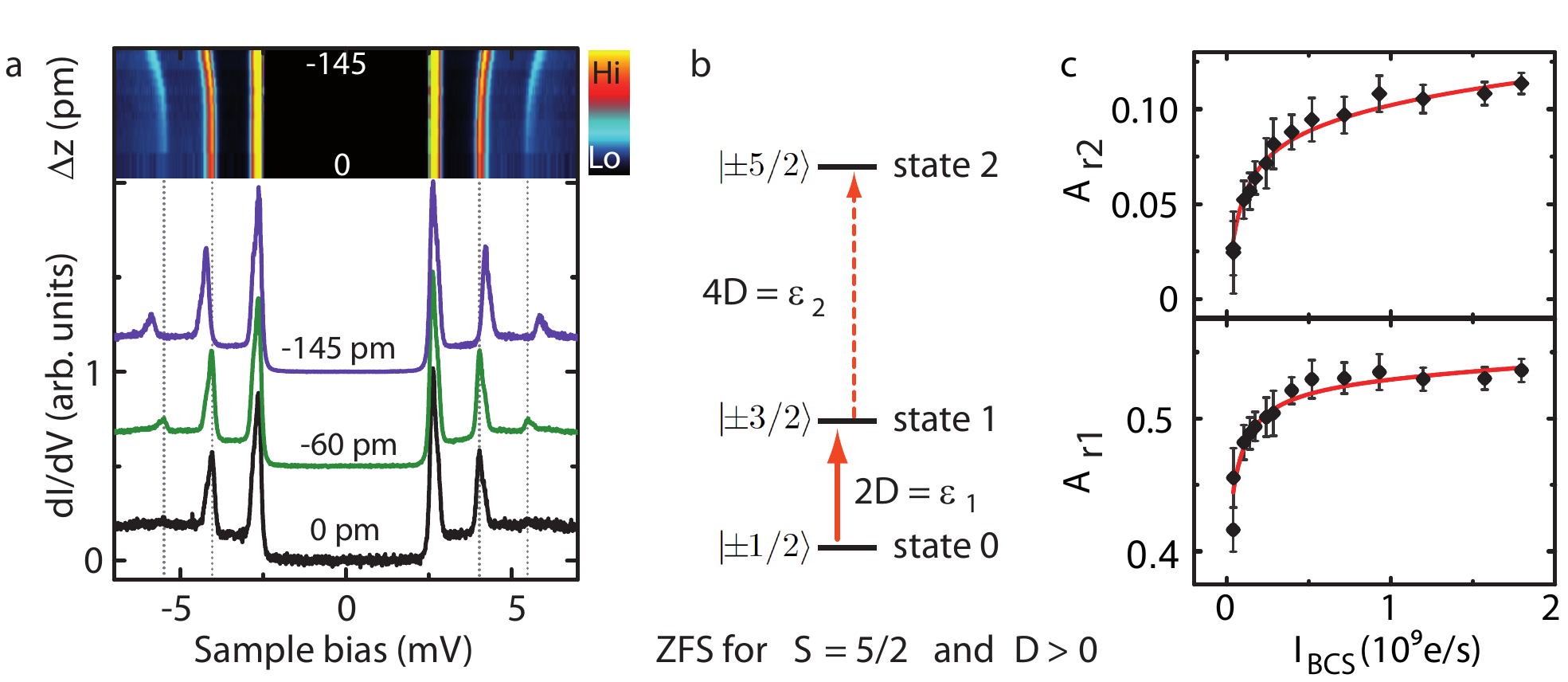}
\caption{Long lived excited spin states probed by distance-dependent excitation spectra.
(a) \textit{dI/dV} spectra acquired at different tip-sample distances reveal: (i) the intensity of both inelastic excitations increases, and (ii) the excitation energies shift to higher energies with decreasing distance. The top panel shows a 2D intensity plot of a series of distant-dependent \textit{dI/dV} spectra. All spectra are normalized to unity at the energy of the quasi-particle peaks and offset for clarity ($I=200$~pA, $V=50$~mV; $\Delta z$ ranging from 0 to -145~pm).
(b) Scheme of the Zero-Field-Splitting for $S=\frac{5}{2}$ with in-plane anisotropy ($D>0$). Due to conservation of angular momentum a tunneling electron can only change the spin state by $\Delta M_S= \pm 1$ or $0$~\cite{hirji06, hirji07}. 
 (c) Relative amplitude $A_{r1}$ ($A_{r2}$) of the first (second) excitation resonance as a function of $I_{BCS}$, the current integral of the BCS peaks. The error bars are determined via error propagation from the uncertainties of the respective amplitudes.
The lifetime of the excited state is fit to the asymptotic increase as described in the manuscript (red line) [Fit parameters for $A_{r1}$: $\tau_1 = 12\pm3$~ns; $P = 0.39\pm0.02$]. 
For $A_{r2}$, the fit is described in the Supplementary Information ($\tau_1 = 12.3$~ns; $\tau_{2\rightarrow 0}=100$~ns;  free fit parameters: $\tau_{2\rightarrow 1} =400\pm 100$~ps;  $P=0.25\pm0.02$; $\chi ^2 = 0.16$.)}
\label{Fig2}
\end{figure}

A careful analysis of the excitation spectra of Fe-OEP-Cl reveals signatures of a second excitation at an energy $\varepsilon_2=2.8$~meV, i.e., at $2\times\varepsilon_1$ (dashed circle in Fig.~\ref{Fig1}c). From these two excitations we deduce the $S=\frac{5}{2}$ spin state of the Fe-OEP-Cl molecule. Only this multiplet of Fe$^{3+}$ splits in the presence of axial magnetic anisotropy into 3 doublets, $M_S=\pm \frac{1}{2}$; $\pm \frac{3}{2}$; and $\pm \frac{5}{2}$ (Fig.~\ref{Fig2}b), separated by energies of $2D$ and $4D$~\cite{gatteschi06}. The observation of two inelastic excitations with $\varepsilon_2=2\times\varepsilon_1$ is only possible if the Kramer doublet with $M_S=\pm\frac{1}{2}$ is the ground state of the multiplet (as will be presented in the following). The resulting in-plane magnetic anisotropy with $D=0.7$~meV is in sign and magnitude similar to the known values for Fe-OEP-Cl in bulk samples~\cite{nishio01}, indicating again the minor interaction between the molecular spin and the Pb(111) substrate.

Although a spin $S=\frac{5}{2}$ allows for two zero field excitations with $\Delta M_S= \pm 1$, the higher-energy one ($\pm \frac{3}{2}\rightarrow \pm \frac{5}{2}$) is only possible if the intermediate state $M_S=\pm \frac{3}{2}$ is populated with a finite probability. At the low temperature of our experiment, the thermal occupation of the intermediate state is negligible, but it can be populated through preceding inelastic tunneling events. In fact, we observe that the second excitation emerges in the spectra when the tip is brought closer and larger tunneling currents are applied (Fig.~\ref{Fig2}a), which is a fingerprint of spin pumping into higher-lying excited states \cite{loth10Nat}.

To monitor how the tunneling current enables the population of excited spin states we plot in Figure~\ref{Fig2}c the current dependence of the relative amplitude of the two inelastic peaks, $A_{r1}$ and $A_{r2}$ (see \textit{Methods}).
 An asymptotic increase of the relative amplitudes with the current is observed for both excitations. For the first one, $A_{r1}$ increases about $20\%$ and saturates for currents larger than $0.6 \times 10^9$ electrons per second ($\approx$ 0.1~nA). The second excitation is hardly detectable at low tunneling current values, but when it appears, $A_{r2}$ increases rapidly with the current $I$. For the smallest current values, the mean time between tunneling electrons is much larger than the excitation lifetime $\tau_1$. Spin excitation from the ground state to state 1 is the only possible inelastic event. The pronounced increase of both inelastic signals with current indicates the activation of additional inelastic processes due to long spin excitation lifetimes. When the inelastic tunneling rate 
equals the natural decay constant of this excitation $\lambda_1=1/\tau_1$, tunneling electrons may find the molecule still populating state 1 and contribute to the spin relaxation by absorbing the excitation energy. This inelastic decay channel is open to all electrons and, hence, causes an increase of $A_{r1}$. The saturation of $A_{r1}$ for currents above 0.1~nA denotes a stationary non-equilibrium state, with transitions between ground and first excited spin state being solely driven by the inelastic current. A consequence of this is a finite population of the first excited state (state 1), which also enables excitations to the second excited state (state 2) ~\cite{loth10Nat}. This is reflected in the appearance of the inelastic signal at $\varepsilon_2 = 2 \times \varepsilon_1$. This scenario corroborates our earlier assignment of $M_S = \pm \frac{1}{2}$ being the groundstate, i.e. $D>0$.

To obtain a precise value for the natural lifetime $\tau_{1}=1/\lambda_1$ of state 1 ($|M_S| = \frac{3}{2}$), we set up rate equations accounting for changes of the ground and excited state occupation ($N_{0}$ and $N_1$). The excitation process is driven by tunneling electrons above the threshold energy $eV \ge2\Delta +\varepsilon_1$, and depends on the inelastic  transition probability $P$, and the elastic current through the BCS peak $I_{BCS}$ (a detailed explanation is included in the Supplementary Information). 
The relaxation process contains, besides the spontaneous decay with constant $\lambda_{1}$, an electron induced inelastic deexcitation.
Since the timescale of our experiment is long compared to the tunneling frequency, we measure a stationary state of the occupations $N_{0}$ and $N_1$, which depends on the tunneling current. From the rate equations (see Supplementary Information), we obtain that the relative amplitude $A_{r1}$ depends on the elastic current $I_{BCS}$ as:
\begin{equation}
A_{r1} = P ~ \frac{(I_{BCS}/e) P (2+2\varepsilon_{1}/\Delta) + \lambda_{1} }{ (I_{BCS}/e) P (2+\varepsilon_{1}/\Delta) + \lambda_{1} }~.\label{Amp}
\end{equation}
 We use equation~(\ref{Amp}) to fit the dependence of $A_{r1}$ with current in Fig.~\ref{Fig2}c and obtain that the lifetime of the first excited state is $\tau_{1} = 1/\lambda_{1} = 12\pm3$~ns, in agreement with the appearance of the second excitation peak for currents in the order of $10^8 e/s$. 
 This value is larger by orders of magnitude than typical spin excitation lifetimes of magnetic atoms on metal substrates, where $\tau$ typically spans up to a few hundred femtoseconds \cite{balashov09PRL,chilianPRB11,khajetPRL11}. Furthermore, it is still larger by more than one order of magnitude than the lifetime of single atoms on top of thin decoupling layers like CuO, BN or Cu$_2$N \cite{tsukahara09,kahleNanoL11,loth10Nat}, which succeeded to extend spin lifetimes up to hundreds of picoseconds~\cite{loth10Nat}. 

For currents larger than $0.1$ nA, the population of the $M_S=\pm \frac{3}{2}$ state increases to about 30\% (see Fig.~S6), allowing tunneling electrons to excite a second transition to the $M_S=\pm \frac{5}{2}$ state (state 2) when their energy reaches the $4D$ threshold value ($\approx$ 2.8 meV). 
The increase of the relative amplitude $A_{r2}$ from zero to $\sim 0.12$ with current is essentially a consequence of the increase of the population of the intermediate state 1, which enables the second excitation 
and corresponds roughly to the product of the stationary, non-equilibrium population of state 1 times the transition probability $P$. 
The fit of $A_{r2}$ as described in the Supplementary Information confirms the lifetime $\tau_1$ of the order of 10~ns. It further hints at a lifetime $\tau_2$ being significantly shorter than $\tau_1$.

To explain the long lifetime of state 1, we note that the most efficient spin relaxation channel on metal surfaces consists of spin scattering with conduction electrons and the creation of electron-hole pairs. On the lead surface, the superconducting energy gap around $E_F$ amounts to $2\Delta$ = 2.7~meV, which is larger than the energy of the first excited state $\varepsilon_1$ = 1.4~meV. Hence, the absorption of the spin excitation energy by electron-hole-pair creation is blocked. The absence of the most efficient relaxation channel extends the state's lifetime. Only other less efficient relaxation channels, such as direct and indirect spin-phonon coupling, can release energy by excitations in the phonon band of the substrate \cite{leuenberger99}. The importance of the energy gap of the substrate is further reflected by the hint that the lifetime of the second excited state is considerably shorter. For this excitation $\varepsilon_2 > 2\Delta$ and it can thus decay directly into the substrate by electron-hole-pair excitations. Such fast excitation decay would also be the case for the first excited state if Fe-OEP-Cl were not adsorbed on superconducting Pb(111), but on a normal metal substrate with similar adsorption properties, as Au(111) (see section~V of the Supplementary Information). On Au(111) no sign of a second excitation is observed for Fe-OEP-Cl, allowing us to  determine an upper limit of 400~ps for the lifetime of $\tau_{1}$. This finding underlines the importance of the superconducting state for the protection of the excited spin state 1.

\begin{figure}[h]
\includegraphics[width=0.96\textwidth,clip=]{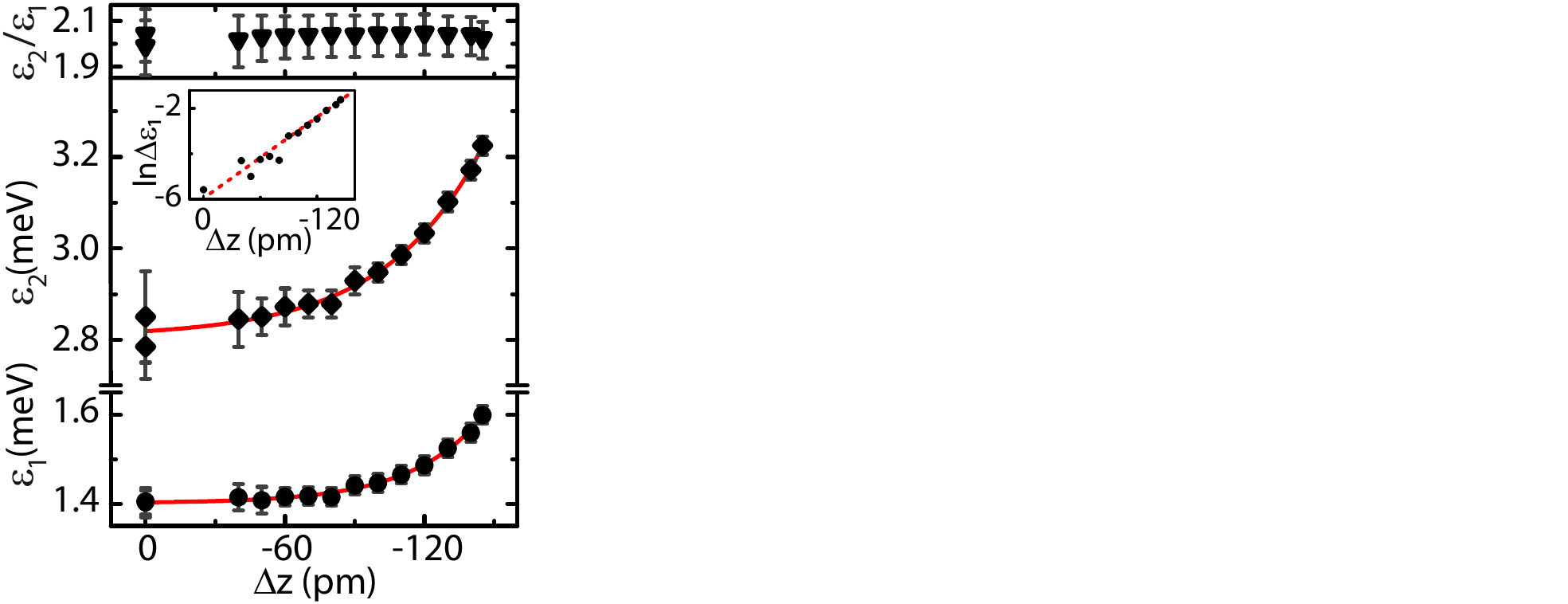}
\caption{Changing the magnetic anisotropy with the STM tip. The excitation energies of the first (bottom) and second (middle) excitation, and their ratio $\frac{\varepsilon_1}{\varepsilon_2}$ are drawn as a function of the relative tip-sample distance $\Delta z$. The error bars are determined via error propagation from the uncertainties of the respective energy position determination.
An exponential fit yields zero-current excitation energies of $\varepsilon_1^0=1.40\pm0.01$~meV, and $\varepsilon_2^0=2.81\pm0.01$~meV. The linear slope in the logarithmic plot (see inset) of the excitation energy difference $\Delta \varepsilon_1 = \varepsilon_1 (\Delta z) - \varepsilon_1^0$ unveils the influence of the increased wave function overlap of tip and molecule on the magnetic anisotropy (\textit{dashed} line as guide for the eye).  }
\label{Energy}
\end{figure}

A remarkable effect observed in Fig.~\ref{Fig2} is the monotonous increase of both spin excitation energies with a decrease in tip--sample distance. Figure~\ref{Energy} shows that the inelastic onsets $\varepsilon_1$ and $\varepsilon_2$ grow exponentially with displacement of the tip, while their ratio amounts to $\frac{\varepsilon_1}{\varepsilon_2}=2$ at every tip position. The constant ratio of 2 indicates that the energy shifts are caused by the continuous increase of the anisotropy $D$ as the STM tip is approached towards the Cl ion, due to an exponentially increasing wave function overlap. $D$ amounts to $0.70 \pm 0.01$~meV at the limit of zero current, \textit{i.e.}, in the absence of the tip, and increases to $0.80 \pm 0.02$~meV at the closest possible position. The gradual increase of $\approx 15~\%$ in the magnetic anisotropy is induced by the proximity of the STM tip to the paramagnetic species and is probably a consequence of the forces exerted to the molecule, which scale exponentially with the tip--molecule distance~\cite{ternesPRL11}. Before the forces become too strong and the chlorine ion is detached, mechanical deformations of the molecule lead to a change in the crystal field of the Fe$^{3+}$ core and hence its magnetic anisotropy. This effect exemplifies the tunability of anisotropy of single atoms by reversible changes in their atomic-scale surrounding \cite{parksScience10}, with the changes being strongest in rather flexible environments such as coordinatively bonded metal-organic complexes.  

Traditionally, atomic magnetism uses semiconducting substrates or insulating layers to extend magnetic excitation lifetimes to levels that allow manipulation of spins. Our results show that the combination of a passive organic ligand and a superconducting substrate preserve magnetic states and spin excitations from decaying for several nanoseconds. 
This time-scale would be long enough for quantum information processing in multi-center molecular magnets~\cite{Leuenberger01}, and for electrical spin pumping and reading~\cite{loth10Nat}.

\section{Methods}
Our experiments were carried out in a \textsc{Specs} JT-STM, an ultra-high vacuum scanning tunneling microscope operating at a base temperature of 1.2 K. 
Spectra of the differential conductance $dI/dV(V)$ were acquired under open-feedback conditions with standard lock-in technique using a modulation frequency of $f=912$~Hz and an amplitude of $V_{rms}=30 - 50~\mu$V. 
The Pb(111) surface (critical temperature $T_c=7.2$~K) was cleaned by repeated sputter/anneal cycles until a clean, superconducting surface was obtained.
The Pb tip was prepared by indenting the tip into the Pb surface while applying a voltage of 100 V. To check the quality of the as prepared tips we record $dI/dV$ spectra on the bare Pb(111) surface at 4.8~K. At this temperature, thermal excitations of the quasi-particles across the superconducting energy gap lead to a finite number of hole-like states at $-\Delta$ and electron-like states at $+\Delta$. If the superconducting gaps of tip and sample are of same width, \textit{i.e.}, $\Delta_{tip} = \Delta_{sample}$, this finite state occupation results in a small conductance peak exactly at zero bias \cite{franke11}. If the gaps are of different width, \textit{i.e.}, $\Delta_{tip} \neq \Delta_{sample}$, we find peaks at $eV=\Delta_{tip}-\Delta_{sample}$ and $eV=-(\Delta_{tip}-\Delta_{sample})$. Throughout the experiment we only used tips that fulfilled $\Delta_{tip}=\Delta_{sample}=\Delta$.
The superconducting state of the tip leads to an increase in energy resolution beyond the intrinsic Fermi-Dirac broadening of a normal metal tip \cite{franke11}, because the substrate's density of states  is  sampled with the sharp quasi-particle peaks of the tip.

Fe-OEP-Cl was sublimated from a crucible at 490~K onto the clean Pb(111) surface held at 120~K. To enhance self-assembly into ordered domains, the sample was subsequently annealed to 240~K for 180~s, prior to cooling down and transferred into the STM.
Ordered monolayer islands of quasi-hexagonal structure can be identified, with the ethyl-groups clearly visible in the STM images (Fig. 1a). About 30$ \%$ of the molecules show a bright protrusion in their center. Annealing to higher temperatures after deposition reduces the number of protrusions.
We can thus identify the protrusion as the central Cl ligand which is present when the molecules are evaporated from the powder. We can exclude any impurities adsorbed from the background pressure by a series of different preparations. The fraction of chlorinated molecules does not depend 
on the time for which the prepared sample is kept in the preparation chamber. The only observable influence is an elevated annealing temperature after molecule deposition. Reference~\cite{wende07} reports complete dechlorination upon adsorption on thin Ni and Co layers at 300~K, while Fe-OEP-Cl deposited on Au(111) at $240$~K, retains the Cl ligand almost completely~\cite{bheinrich13}. We focus our study on the molecules retaining their central chlorine ligand.
 
For the analysis of the excitation lifetime, $A_{r1}$ and $A_{r2}$ are quantified as the relative amplitude of the peaks appearing at the two inelastic onsets with respect to the amplitude of the BCS peak. They are defined as the ratio of amplitudes of the inelastic peak and the BCS peak in the $dI/dV$ spectra: $A_{r1} = \frac{A_{1}}{A_{BCS}}$ and $A_{r2} = \frac{A_{2}}{A_{BCS}}$, for the first and second excitation, respectively. Due to the peaked nature of the superconducting density of states, this corresponds to the commonly employed measure of the increase of the differential conductance at the threshold of the excitation.
$A_{1}$ ($A_{2}$), and $A_{BCS}$ are determined as the mean amplitudes of positive and negative bias side of the first (second) excitation and of the BCS peak. For details see the Supplementary Information.

\section{Acknowledgments}
We thank Piet Brouwer, Felix von Oppen, Nicol\'as Lorente and Markus Ternes for fruitful discussions. Financial support by the Deutsche Forschungsgemeinschaft through Sfb 658 and by the focus area Nanoscale of Freie Universit\"at Berlin is gratefully acknowledged.

\section{Author contributions}
B.W.H., J.I.P. and K.J.F. designed the experiments. B.W.H. and L.B. performed the
experiment. All authors discussed the data analysis and the results. B.W.H., J.I.P. and
K.J.F. co-wrote the paper.

\section{Competing Financial Interests}
The authors declare no competing financial interests.

\newpage
\Large{Supplementary Information:}\normalsize
\\
\\

\section{I. First excitation}
\label{1Exc}
To determine the lifetime of the excited spin states, we analyze the amplitude of the inelastic signals in the $dI/dV(V)$ spectra. 
As shown in the manuscript, the ground state (state $0$) corresponds to the spin eigenstates $|M_s | = \frac{1}{2}$, the first excited state ($1$) to  $|M_s | = \frac{3}{2}$, and the second excited state ($2$) to $|M_s | = \frac{5}{2}$.

To simplify the analysis of the inelastic intensities, we first restrict ourselves to the threshold energy of the first excitation, \textit{i.e.}, at $|eV |= 2\Delta +\varepsilon_1$.  Hence, the second transition is not yet active.

\subsection{Rate equations}
\label{rates}

We start by writing down the rate equations for changes in the occupation $N_0$ and $N_1$ of the ground and first excited state, respectively:\\

$\frac{dN_{0}}{dt}=R_{1 \rightarrow 0}N_{1}-R_{0 \rightarrow 1}N_{0} = 0$~, and

 $\frac{dN_{1}}{dt}=R_{0 \rightarrow 1}N_{0}-R_{1 \rightarrow 0}N_{1} = 0$~,\\

\noindent where $R_{i \rightarrow j}$ is the rate constant of the transition from state $i$ to state $j$ [$i,j \in \{0,1\}$]. As our measurements describe a stationary state, any changes are set to zero.

The rate constants are defined by a spontaneous decay constant $\lambda_1$ (short for ``{$\lambda_{1 \rightarrow 0}$}''), and a current-induced term. The later is determined by the transition probability  $P$ for excitation/deexcitation times the rate of tunneling electrons with sufficient energy for the inelastic excitation. The phenomenological factor $P$, which we call transition probability, may here also include multiplicative factors such as geometrical effects or tip position. The special case of the superconductor-superconductor tunneling junction allows us to derive the rate of tunneling electrons as a function of the (elastic) tunneling current through the BCS-like quasi-particle peak ($I_{BCS}$), \textit{i.e.}, the current integral over the peaks in the \textit{dI/dV} spectra. $I_{BCS}$ is directly determined from the \textit{I(V)} spectra as shown in Fig.~\ref{I_supp}a. We assume the transition matrix element in the integral for the tunnel current to be independent of energy in the small bias range of our measurements. If the transition probability $P$ were one, the BCS peak would appear with the same intensity at the threshold energy, because the excitation process connects the BCS peaks in the DoS of tip and sample.  This would then correspond to an inelastic tunneling current of the same magnitude as $I_{BCS}$. Hence, the excitation rate can be related to this particular current multiplied by the transition probability $P$ and divided by the elementary charge $e$: $R_{0 \rightarrow 1} = P~ I_{BCS} /e$~. \\

\begin{SCfigure}
  \includegraphics[width=0.50\textwidth,clip=]{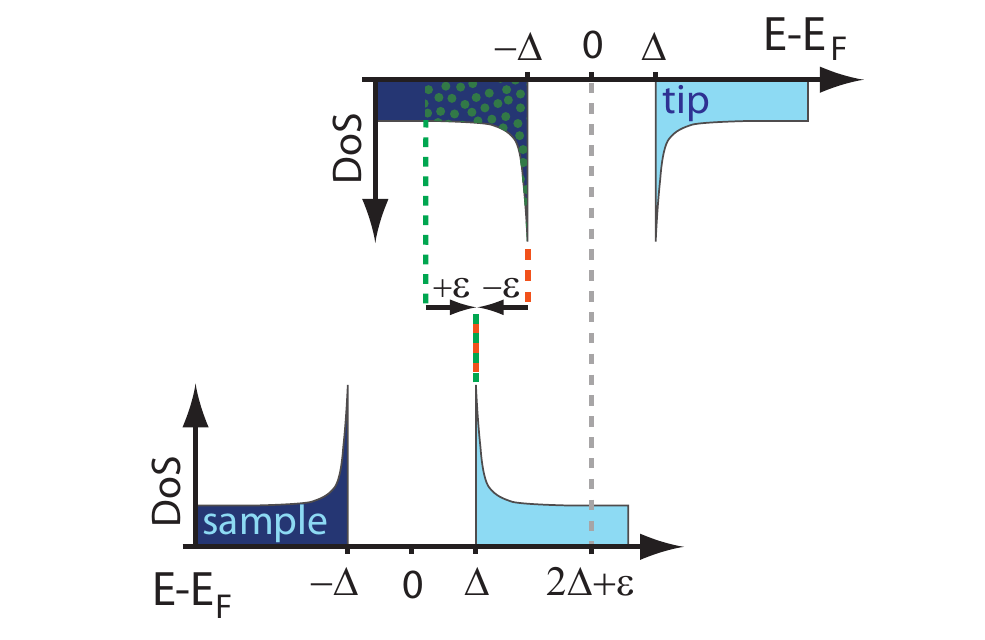}
  \caption{Scheme of the inelastic tunneling at the threshold energy of $|eV| = 2\Delta + \varepsilon$. The number of electrons with the right energy for the deexcitation (green) is larger by $2\varepsilon$ compared to the excitation.}
\label{S_supp}
\end{SCfigure}

The deexcitation can not only be induced by the electrons with the threshold energy, but also by electrons with lower energy (see Fig.~\ref{S_supp}). The number of electrons, which are able to deexcitate the spin state, is given by the energy window defined by the sample bias and the energy gain due to deexcitation: $eV+\varepsilon_1=2\Delta+2\varepsilon_1$. To account for this larger number of electrons, we introduce a factor $f$, which describes the increase in available electrons with respect to $I_{BCS}$: $R_{1 \rightarrow 0} = \lambda_{1} +  f~P~I_{BCS}/e$~.\\

As shown in Figure~\ref{I_supp}, in the superconductor-superconductor junction, the elastic tunneling current follows -- in a good approximation -- an ohmic behaviour with
$I \propto eV$ for $|eV| \ge 2\Delta$. Hence, the number of electrons capable for tunneling is proportional to the applied bias. The factor $f$ can be defined from the ratio of electrons in the energy window for the deexcitation and the elastic tunneling at the BCS peaks:
\begin{equation*}
f=\frac{2\Delta +2\varepsilon_1}{2\Delta}=1+\varepsilon_1/\Delta~.
\end{equation*}

\subsection{State occupation}

\begin{figure}
  \includegraphics[width=0.98\textwidth,clip=]{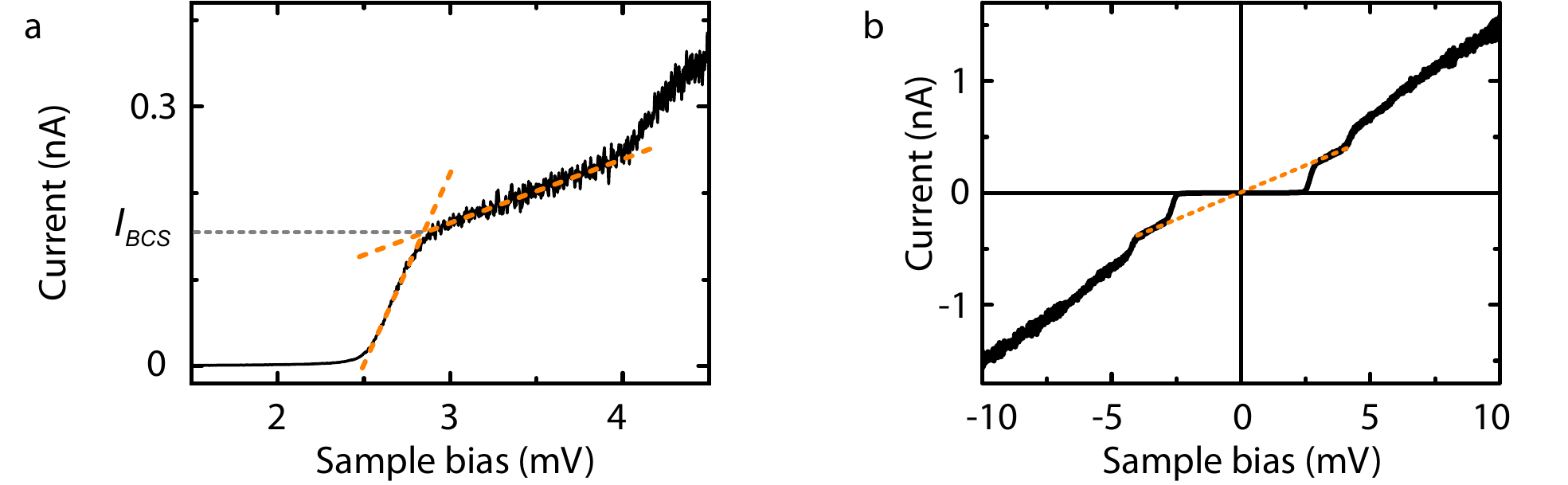}
  \caption{$I-V$ characteristic of the superconductor-superconductor tunnel junction.
a) We determine $I_{BCS}$, the current due to tunneling through the BCS peaks, as the intersection value of two straight lines approximating the curve as shown in the graph. This corresponds to the integral of the BCS peaks in the $dI/dV$ spectra. 
 b) In the limit of low biases and with $|eV| \ge 2\Delta$, the elastic current is proportional to the sample bias ($I_{elastic}= \frac{V}{R}$, with $R$ being the nominal (\textit{elastic}) resistance of the junction).}
\label{I_supp}
\end{figure}

The rate equations can now be linked to the time-average occupation of state 0 and 1.\\
We can set the total occupation to one:
\begin{equation}
N_{0}+N_{1} = 1~.
\label{occupation}
\end{equation}

\noindent Introducing the rate equations and rate constants from section~III into eq.~(\ref{occupation}), we can write:
\begin{equation}
 \frac{N_{1}}{1 - N_{1}} = \frac{ P }{(1+\varepsilon_1/\Delta)~   P + \frac{ \lambda_{1}}{I_{BCS}e^{-1}}}~.
 \label{N1}
\end{equation} \\

\subsection{Inelastic intensity}
To obtain the state occupation and lifetime from our experimental data, we determine the relative amplitude of the inelastic peak, which is defined as the amplitude ratio of the  inelastic and the BCS peak:
$A_{r1} =A_{1}/A_{BCS}$. $A_{BCS}$ is the absolute amplitude of the BCS peak in the $dI/dV$ spectra, averaged over the peaks at positive and negative bias, $A_{1}$ the amplitude of the first excitation peak, \textit{i.e.}, the difference of the $dI/dV$ value at the peak and the background value before the peak, again averaged over positive and negative bias.
This corresponds to the commonly employed measure of the increase of the differential conductance at the threshold of the excitation.
It is equal to the ratio of current increase at the inelastic excitation and the BCS peak, respectively:
$A_{r1} = I_1/I_{BCS} = I_{r1}$, where $I_{1}$ is the current integral over the excitation peak of the first excitation (similar to $I_{BCS}$ for the BCS peak).
The total inelastic tunneling current consists of an excitation and a deexcitation current.
Hence,\\

$I_{1} = I_{BCS} ~  P \left( N_{0} +N_{1}~ (1+\varepsilon_1/\Delta)\right)$.\\

This yields\\

$A_{r1}=I_{r1} = P \left( N_{0} +N_{1}~ (1+\varepsilon_1/\Delta)\right)$.\\

Using again eq.~(\ref{occupation}), this is reduced to

\begin{equation}
\frac{P (1+\varepsilon_1/\Delta)-A_{r1}} {P~\varepsilon_1/\Delta } = N_{0}~.
\label{Inel}
\end{equation}\\

We combine eq.~(\ref{N1}) and eq.~(\ref{Inel}) to obtain:\\
\begin{equation}
A_{r1}  =  P ~ \frac{\frac{I_{BCS}}{e}~ P~(2+2\varepsilon_1/\Delta) +  \lambda_{1} }{ \frac{I_{BCS}}{e}~P ~( 2+\varepsilon_1/\Delta ) + \lambda_{1}}~.
\end{equation}\\

\begin{SCfigure}
  \includegraphics[width=0.50\textwidth,clip=]{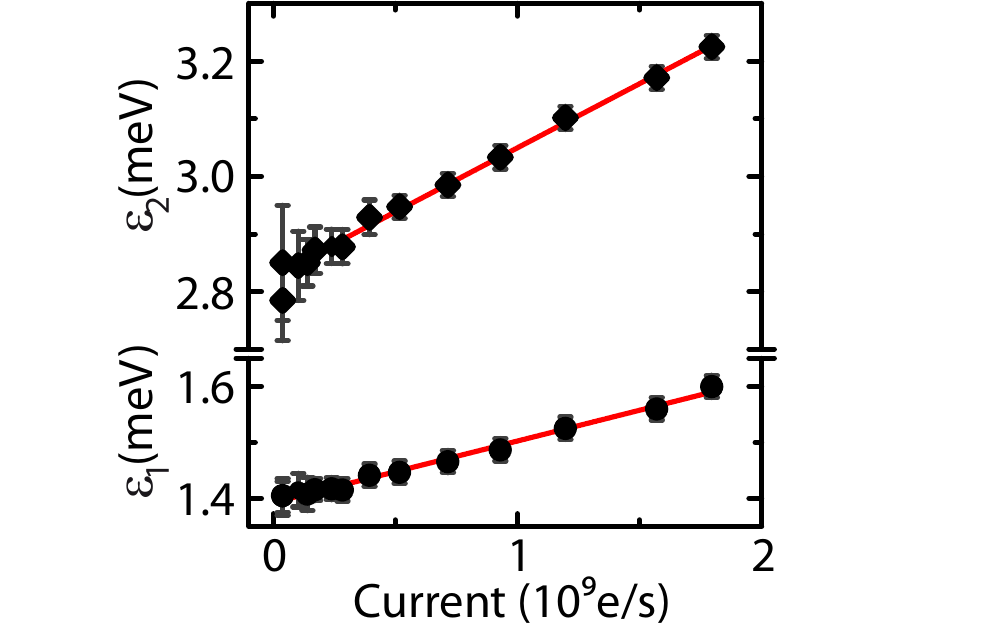}
  \caption{Linear dependency of the excitation energies $\varepsilon_{1}$ and $\varepsilon_{2}$ with the tunneling current through the BCS peaks $I_{BCS}$}.
\label{E_supp}
\end{SCfigure}
This formula provides the direct relation between the excitation  amplitude and the lifetime of the first excited state. For small currents the amplitude only depends on the probability $P$. For higher currents however, when the tunneling rate is much higher than the decay rate, the amplitude increases by a factor of $\frac{(2+2\varepsilon_{1}/\Delta)}{(2+\varepsilon_{1}/\Delta)}$.\\
To account for the change in the excitation energy  with the tip--sample distance (see Fig.~3 in the manuscript), we fit the linear increase of $\varepsilon_{1}$ and $\varepsilon_{2}$ with the current as shown in Fig.~\ref{E_supp}. The obtained slope is included into the fitting routine of the lifetime for completeness.

\section{II. Second excitation}
\label{SecE2}
To determine the lifetime of the second excitation, we now analyze the amplitude of the inelastic signal of the second excitation. The sample bias is given by $eV=2\Delta+\varepsilon_{2}$.
We start by writing down a new set of rate equations, taking  into account all possible transitions:\\

$\frac{dN_{0}}{dt}=R_{1 \rightarrow 0}N_{1}-R_{0 \rightarrow 1}N_{0} + R_{2 \rightarrow 0}N_{2} = 0$ ,

$\frac{dN_{1}}{dt}=R_{0 \rightarrow 1}N_{0}-R_{1 \rightarrow 0}N_{1}+R_{2 \rightarrow 1}N_{2}-R_{1 \rightarrow 2}N_{1} = 0$ ,

$\frac{dN_{2}}{dt}=R_{1 \rightarrow 2}N_{1}-R_{2 \rightarrow 1}N_{2} - R_{2 \rightarrow 0}N_{2} = 0$~.\\

\noindent The different rate constants are given as follows: \\

$R_{0 \rightarrow 1} = f_{0 \rightarrow 1}  P {I_{BCS}}/{e}$~,

$R_{1 \rightarrow 0} = \lambda_{1} +  f_{1 \rightarrow 0} P {I_{BCS}}/{e}$~,

$R_{1 \rightarrow 2} =  P {I_{BCS}}/{e}$~,

$R_{2 \rightarrow 1} = \lambda_{2 \rightarrow 1 }  + f_{2 \rightarrow 1}  P {I_{BCS}}/{e}$~, and

$R_{2 \rightarrow 0} = \lambda_{2 \rightarrow 0}$ .\\

\noindent As for the first excitation in section~III, we have to consider different numbers of electrons for the current induced processes due to the different energy windows accessible.
The additional factors $f_{i \rightarrow j}$ read here:\\

$f_{2\rightarrow1}=\frac{2\Delta +2\varepsilon_2}{2\Delta}=  1+\frac{\varepsilon_{2}}{\Delta}$~,\\

$f_{0\rightarrow1}= \frac{2\Delta + \varepsilon_2 - \varepsilon_1}{2\Delta}= 1+\frac{\varepsilon_{2}-\varepsilon_{1}}{2\Delta}$~, and\\

$f_{1\rightarrow0}=\frac{2\Delta + \varepsilon_2 + \varepsilon_1}{2\Delta}=  1+\frac{\varepsilon_{2}+\varepsilon_{1}}{2\Delta}$~.\\

\noindent It is noteworthy that also for the first excitation process the number of electrons capable of excitation is increased compared to the second excitation. The applied sample bias is larger than the threshold of the first excitation:
$eV=2\Delta+\varepsilon_{2} > 2\Delta+\varepsilon_{1}$.  So more electrons can contribute.

As shown above in detail for the first excited state, we now similarly deduce the relative amplitude $A_{r2}$ as follows:

\begin{small}
\begin{multline*}
\label{Ar2}
 \left.A_{r2}  = \right.\\
\left.\frac{
P^{2} \frac{I_{BCS}}{e} \left(1+\frac{\varepsilon_{2}-\varepsilon_{1}}{2\Delta}\right)
 \left[2 P  \frac{I_{BCS}}{e}  \left(1+\frac{\varepsilon_{2}}{\Delta}\right)+  {\lambda_{2 \rightarrow 1}} + \lambda_{2 \rightarrow 0}\right] }
{\lambda_1 \left(\lambda_{2 \rightarrow 1}+\lambda_{2 \rightarrow 0}\right)
+ P \frac{I_{BCS}}{e} \left[
\lambda_1\left(1+\frac{\varepsilon_{2}}{\Delta}\right)
+\lambda _{2 \rightarrow 1}\left(2+\frac{\varepsilon_{2}}{\Delta}\right)
+  \lambda_{2\rightarrow 0} \left(3+\frac{\varepsilon_{2}}{\Delta}\right)
\right]
+ P^{2} (\frac{I_{BCS}}{e})^2
\left[3+\frac{7\varepsilon_{2}}{2\Delta} -\frac{\varepsilon_{1}}{2\Delta}+\frac{\varepsilon_{2}^{2}}{\Delta^{2}}
\right]}. \right.
\end{multline*}
\end{small}

\noindent This allows for a fitting of the decay constants of the second excited state. The lifetime $\tau_2$ is then calculated as:

\begin{equation*}
\tau_{2}=\frac{1}{\lambda_{2\rightarrow 0}+\lambda_{2\rightarrow 1}}~.
\end{equation*}\\

\section{III. Fitting of the experimental data with rate equations}

With the above deduced formulas the current-dependent relative amplitudes $A_{r1}$ and $A_{r2}$ have been fitted to the experimental data presented in the main manuscript. This results in the spontaneous decay rates $\lambda_1$, $\lambda_{2\rightarrow1}$, and $\lambda_{2\rightarrow0}$, as well as the excitation lifetimes $\tau_1=1/\lambda_1$, and $\tau_2=1/(\lambda_{2\rightarrow1}+\lambda_{2\rightarrow0})$, respectively.

\subsubsection{First excitation:} We use eq. (4) to fit the increase and saturation of the relative amplitude $A_{r1}$ of the first excitation peak with I$_{BCS}$. The fit is robust against initiation with different parameters and converges into a well defined set of parameters: the transition probability $P=0.39\pm 0.02$ and the lifetime $\tau_1= 12 \pm 3$~ns. This lifetime is in agreement with the saturation behavior of $A_{r1}$ for currents of some 10$^8$ e/s.
This is the most important outcome of our work, as such long lifetimes are well beyond known values of electron and spin excitation lifetimes on metals. 
We directly test the role of the superconducting gap in section~V, where we compare to Fe-OEP-Cl on Au(111). The upper limit of the lifetime on this metal surface was determined to be 400~ps, \textit{i.e.}, more than one order of magnitude smaller. This underlines the importance of the superconducting state for the long lifetime of the excited spin state 1.
\begin{figure}
  \includegraphics[width=0.98\textwidth]{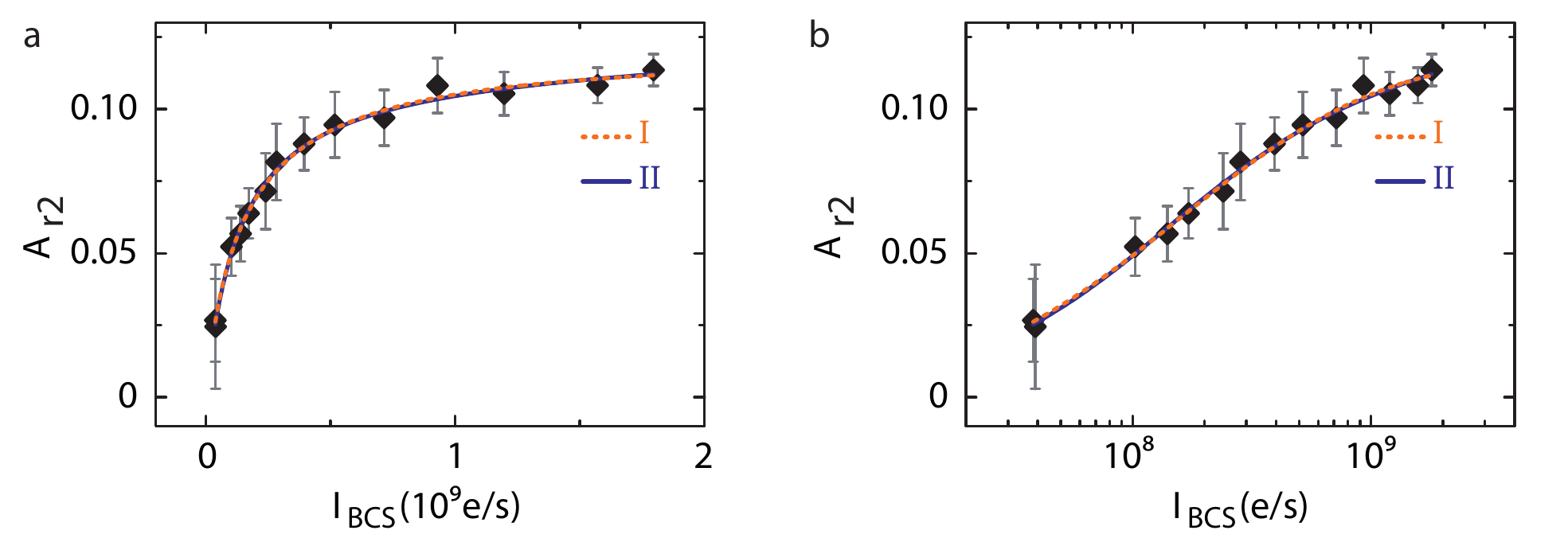}
  \caption{Fit of $A_{r2}$ using the expression obtained in section~II. Two local minima of the error function are obtained. The corresponding curves are labeled I and II in the graph. (a) and (b) show the same fit in linear and logarithmic horizontal scale, respectively. Fit {I}: $\tau_1 =6\pm1$~ns; $\tau_{2\rightarrow 1} =35\pm 30$~ps;  $P=0.31\pm0.02$;  $\chi ^2 = 0.07$. Fit II: $\tau_1 =18\pm 3$~ns; $\tau_{2\rightarrow 1} =7\pm 3$~ns;  $P=0.18\pm0.01$;  $\chi ^2 = 0.08$. $\tau_{2\rightarrow 0}$ was fixed in both fits to 100~ns.\cite{noteS4} }
\label{fit1}
\end{figure}

\begin{figure}
  \includegraphics[width=0.98\textwidth]{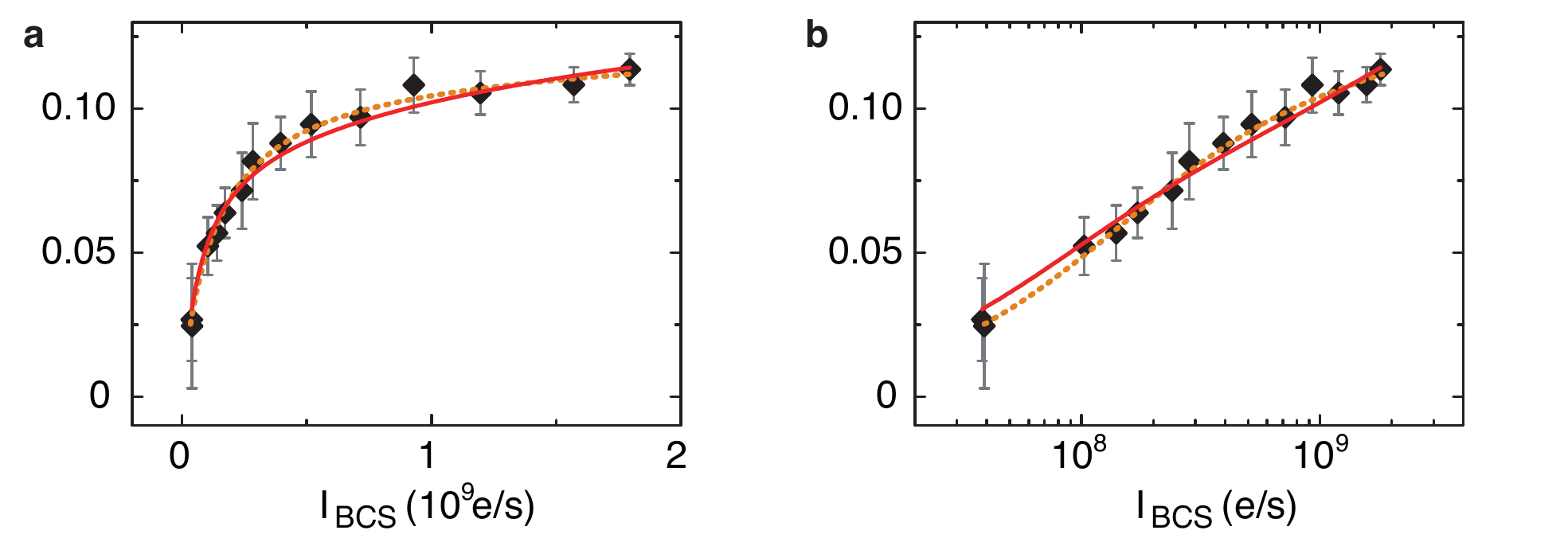}
  \caption{Fit (\textit{full red line}) of $A_{r2}$  using the expression obtained in section~II and fixing  $\tau_{1}$ to the value obtained in the first fit. For comparison the \textit{dashed orange line} shows fit II as presented in Fig.~\ref{fit1}. Fit parameters:  $\tau_{2\rightarrow 1} =400\pm 100$~ps;  $P=0.25\pm0.02$; $\tau_1 = 12.3$~ns (fix); $\tau_{2\rightarrow 0}=100$~ns (fix);\cite{noteS4} $\chi ^2 = 0.16$. }
\label{fit2}
\end{figure}

\subsubsection{Second excitation:}

The second excitation corresponds to transitions between two excited spin states, following from the non-equilibrium occupation of the first excited state. The most important observation is that the amplitude $A_{r2}$ starts at zero and increases with the tunneling current up to $\sim0.12$ with a curvature similar to the plot of $A_{r1}$. This fact is a consequence of the increasing population of the first excited state and its long lifetime $\tau_1$ obtained from the fit of $A_{r1}$.

We fitted $A_{r2}$ using the expression obtained in section~II to obtain a value for the lifetime of the second excitation $\tau_2$. Due to the larger number of fitting parameters this fit yields two local minima of the error function (shown in Fig.~\ref{fit1}). One of the minima (fit I in Fig.~\ref{fit1}) yields $\tau_1 \sim 6$~ns and $\tau_2 \sim 35$~ps, with $P =0.31$. The second minimum (fit II in Fig.~\ref{fit1}) corresponds to the case where both excitations have similar lifetimes ($\tau_1 \sim 18$~ns and $\tau_2 \sim 7$~ns with $P=0.18$) and contribute equally to the increase of $A_{r2}$. Therefore, the curvature of $A_{r2}$ can be similarly reproduced by inducing a sizeable population of either state 1 only, or both, state 1 and state 2, with increasing current. 

To obtain a more reliable estimation of the magnitude of $\tau_2$ we fixed $\tau_1$ to the value obtained from the fit of $A_{r1}$, at 12~ns. This fit yields now a single minimum of the error function, with a lifetime for the second excitation of $\tau_2 \sim 400$~ps, a factor of 30 smaller than $\tau_1$ (see Fig.~\ref{fit2}). 

An outcome of this last fit is that $P=0.25\pm0.02$, which is smaller than the value obtained from the more robust fit to $A_{r1}$. It is known that $P$ may vary depending on the spin excitation.\cite{Gauyacq12} A different inelastic probability for the excitation from state 1 to state 2 is probably the reason for the different values obtained for the parameter $P$ in the fit of $A_{r2}$, which here is used to similarly describe both excitations (0 to 1 and 1 to 2) equally. Therefore, although the fit procedures described here seem to indicate that $\tau_1>>\tau_2$, hence supporting the protecting character of the superconducting surface, we refrain from making explicit quantitative statements about the order of magnitude of $\tau_2$.

\section{IV. Current-dependent occupation \lowercase{of the} spin eigenstates}

\begin{figure}
  \includegraphics[width=0.65\textwidth]{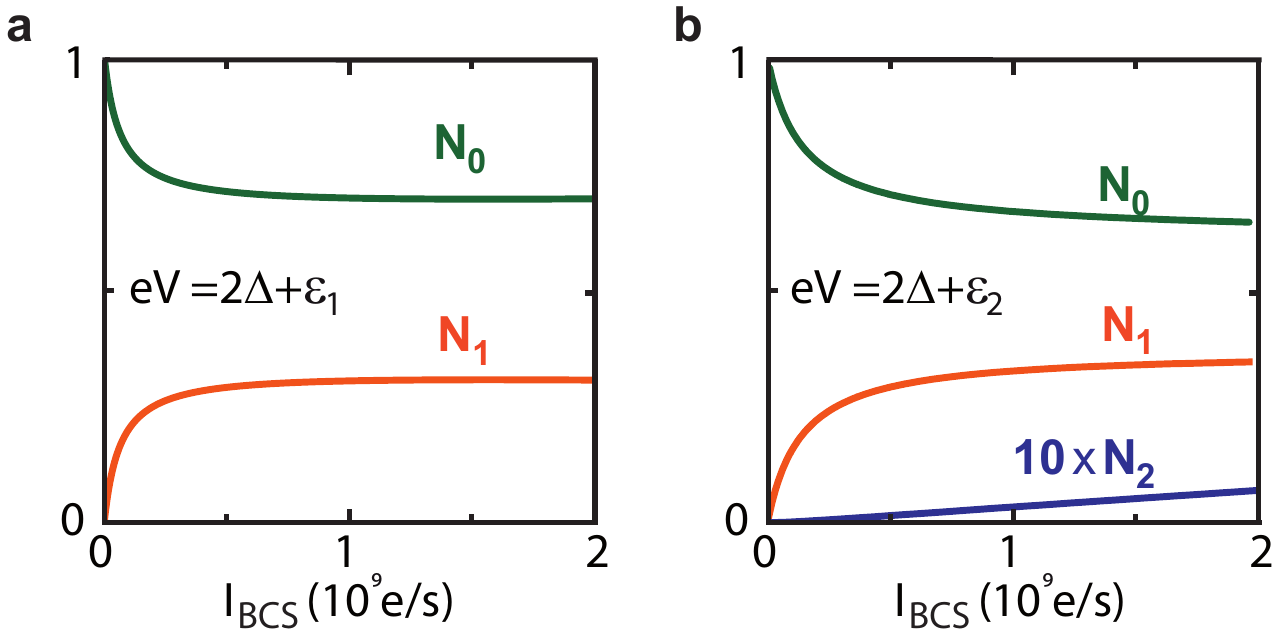}
  \caption{Time-average occupation. N$_{0}$, N$_1$ and N$_{2}$ are calculated from the lifetime of the excited states. a) $eV=2\Delta+\varepsilon_1$. Only N$_0$ and N$_1$ are populated.  Parameters: $\tau_1 =12$~ns; $P=3.9$.
b) $eV=2\Delta+\varepsilon_2$. N$_2$ becomes slightly occupied with increasing currents. Parameters: $\tau_1 =6$~ns; $\tau_{2\rightarrow 1} =35$~ps;  $\tau_{2\rightarrow 0}=100$~ns; $P=0.31$.
We use the parameters extracted from the respective best fits for $A_{r1}$ and $A_{r2}$. }
\label{N_supp}
\end{figure}

With the parameters obtained from the fits, we can plot the occupation N$_{0}$, N$_{1}$, and N$_2$ of the ground, first and second excited state, respectively, as a function of current at the $BCS$ peaks $I_{BCS}$.  We use eq.~(\ref{N1})  together with the parameters extracted from the fit in Fig.~2c in the main paper to calculate N$_{0}$ and N$_{1}$ for a sample bias of $eV=2\Delta+\varepsilon_1$ (Fig.~\ref{N_supp}a). Similarly, the  occupation N$_{0}$, N$_{1}$, and N$_2$ are calculated for $eV=2\Delta+\varepsilon_2$ (Fig.~\ref{N_supp}b). As a result of the different decay rates from first and second excited states, the occupation of N$_{1}$ reaches saturation for roughly $1.5\times 10^9$e/s, while saturation is by far not reached for N$_2$ within experimental accessible conditions.

\section{V. F\lowercase{e}OEP-Cl {on} A\lowercase{u}(111) - adsorption {on a} normal metal}
\label{gold}
\begin{figure}
  \includegraphics[width=0.98\textwidth]{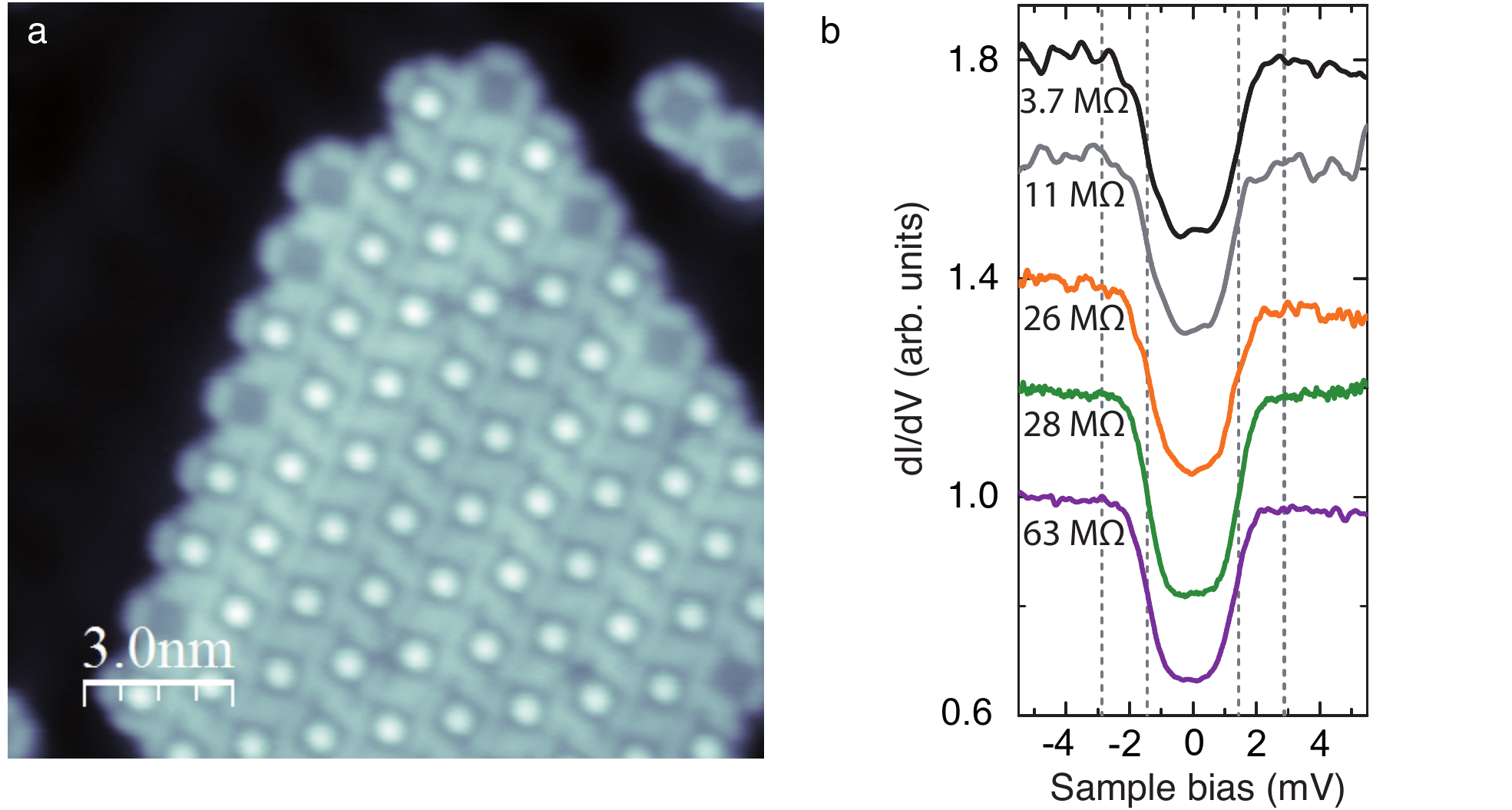}
  \caption{Fe-OEP-Cl on Au(111).
a) Topograph of a mixed island of Fe-OEP-Cl and Fe-OEP on Au(111) [$V=17$~mV, $I=11$~pA]. b) $dI/dV(V)$ spectra acquired in the center of Fe-OEP-Cl molecules adsorbed on Au(111) [nominal junction resistance as noted in the graph, $V_{rms} = 45$ -- $90~\mu$V, $T=2$~K].}
\label{Au_supp}
\end{figure}

To further highlight the importance of the superconducting gap for the long lifetime, we deposit Fe-OEP-Cl on Au(111). This is a non-reactive metal substrate, ensuring little chemical interaction, which could otherwise alter the magnetism of Fe-OEP-Cl. The deposition of the molecules was done as described in the main manuscript for Pb(111). Quasi-hexagonal monolayer islands of Fe-OEP-Cl and Fe-OEP are observed (Fig.~\ref{Au_supp}a), similar to the adsorption on Pb(111).

The only obvious difference with the \textit{on-}Pb(111) case is that on Au(111) the dechlorination process is less frequent at room temperature, and had to be activated by raising the temperature.
As the dechlorination is activated on the surface, this fact hints at a, at least slightly, weaker interaction of Fe-OEP-Cl with the Au(111) substrate. 
Hence, it is reasonable to assume that the spin ground state of Fe-OEP-Cl on Au(111) is also $S=5/2$ with positive anisotropy ($D>0$), as for the free molecule,\cite{wende07} the solid crystal,\cite{nishio01} and the molecule on Pb(111).

Figure~\ref{Au_supp}b shows $dI/dV$ spectra acquired with a Au--covered tip above the center of Fe-OEP-Cl for different junction resistances, \textit{i.e.}, for different tip-sample distances. We observe a step-like increase of the differential conductance at $|eV| = 1.4$~meV due to the opening of an inelastic tunneling channel.\cite{noteS1} This is similar to the first excitation detected in the case of Fe-OEP-Cl on Pb(111) and is a good indicator that neither spin state nor magnetic anisotropy change (within our resolution) due to the adsorption on the different substrates. However, on Au(111) we do not observe any sign of a second excitation at $|eV| = 2.8$~meV (or any other energy), regardless of the junction resistance. It is noteworthy that the lowest resistance measured here is lower than the one measured on Pb(111). This shows that the lifetime of the first excited state is strongly reduced on Au(111) compared to Pb(111). At $R=3.7$~M$\Omega$, the inelastic portion of the current  at $|eV| = 2.8$~meV due to the first excitation amounts to approximately $1.3\times 10^{9}~e\cdot s^{-1}$. This corresponds in average to one inelastic electron every 800~ps.
In a conservative estimation, we can set an upper limit of the lifetime of the first excited state to half this value ($\tau_1 < 400$~ps). If the lifetime were 400 ps, the mean occupation of the first excited state would rise to 0.13 at $|eV|=2.8$~meV.\cite{noteS2} Its absence ensures a lifetime $\tau_1$ shorter than this value for adsorption on the normal metal substrate, \textit{i.e.}, if no gap in the DoS prevents relaxation through electron-hole-pair creation.


\begin{thebibliography}{nnnyys}
\bibitem{noteS1} The DoS of tip and sample are flat in the energy range of interest. Therefore the inelastic signature is a step in the $dI/dV$,\cite{stipe98,heinrich04} not a resonance as in the case of superconducting Pb.
\bibitem{noteS2} We use the same rate equations as presented in Section~SI, but account for the normal metal DoS.  With this we determine the mean occupation of state 1 at the energy of the second excitation ($|eV|=2.8$~meV).
\end{thebibliography}
\end{document}